\documentclass[12pt,preprint]{aastex}
\usepackage{amsmath}
\usepackage{color}
\usepackage{rotating}
\usepackage[normalem]{ulem}

\shorttitle{Radial Flow Pattern of a slow CME }
\shortauthors{Feng et al.}

\begin{document}

\title{Radial Flow Pattern of a slow Coronal Mass Ejection}
\author{Li Feng\altaffilmark{1,2}, Bernd Inhester\altaffilmark{2},
  Weiqun Gan\altaffilmark{1}}
\email{feng@mps.mpg.de; lfeng@pmo.ac.cn}
\altaffiltext{1}{Key Laboratory of Dark Matter and Space Astronomy, 
  Purple Mountain Observatory, Chinese Academy of Sciences,
  210008 Nanjing, China}
\altaffiltext{2}{Max-Planck-Institut f\"{u}r Sonnensystemforschung,
  Justus-von-Liebig-Weg 3, 37077 G\"{o}ttingen, Germany}

\begin{abstract}
  
Height-time plots of the leading edge of coronal mass ejections (CME) have
often been used to study CME kinematics. We propose a new method to analyze
the CME kinematics in more detail by determining the radial mass transport
process throughout the entire CME. Thus our method is able to
estimate not only the speed of the CME front but also the radial flow
speed inside the CME. We have applied the method to a slow CME with an average 
leading edge speed about 480 km s$^{-1}$. In the Lagrangian frame, the speed of 
the individual CME mass elements stay almost constant within 2 and 15 R$_S$, 
the range over which we analyzed the CME. Hence we have no 
evidence of net radial forces
acting on parts of the CME in this range nor of a pile-up of mass ahead of the
CME. We find evidence that the leading edge trajectory obtained by tie-pointing
may gradually lag behind the Lagrangian front-side trajectories derived from our
analysis. Our results also allow a much more precise estimate of the CME
energy. Compared with conventional estimates using the CME total mass and 
leading-edge motion, we find that the latter may overestimate the kinetic 
energy and the gravitational potential energy.
\end{abstract}

\keywords{Sun:coronal mass ejections (CMEs) --- hydrodynamics --- Sun: evolution}

\section{Introduction}

For decades, the kinematics of CMEs have been characterized by the motion of
their leading edge. Since most other parts of the CME show up as either
amorphous or diffuse objects in coronagraph images, few other significant and
recognizable features could reliably be traced in time or associated in stereo
image pairs. Alternatives used in relatively few studies were the column
density barycenter \citep{Mierla:etal:2009} or some distinct 
details of the amorphous core patch sometimes visible in the central region of 
the CME. The plasma of this core could in some cases be traced as continuation 
of the prominence material in EUV images which was ejected along with the CME 
eruption \citep{Mierla:etal:2011}. The leading edge
height-time diagrams are still the main data source to characterize the CME
kinematics. Even though prone to some pointing error, the height diagrams are
repeatedly differentiated to yield velocity and acceleration histories
which then form the base for travel time estimates and speculations about
their dynamics, respectively \citep[e.g., ][]{Mierla:etal:2011, Bein:etal:2011}.

Another key ingredient for estimates on the CME dynamics is the total mass of
the CME. Since the white-light signal in coronagraphs is essentially due to
Thomson scattering at coronal free electrons, a mass estimate from the
coronagraph image brightness seems straight forward. The apparent
mass of a CME, however, does not stay constant with time. Even hours after the
eruption, it continues to rise but more slowly than during the eruption phase. By that time
the leading edge has reached 10~R$_S$ and more. This continuous increase
of the apparent CME mass has been attributed to either a pile-up of solar wind
plasma ahead of the CME or continuous outflow of low coronal plasma from
behind the occulter \citep[e.g., ][]{Bein:etal:2013}.

Kinetic energy estimates of CMEs often simply combine the total mass estimate
with the leading edge velocity \citep[e.g., ][]{Carley:etal:2012}. This
ignores any variation of the flow speed and variations of the mass
distribution inside the CME. \citet{Ventura:etal:2002} derived the distribution 
of the line-of-sight flow
velocity in two CMEs from Doppler-shifts of the O VI line observed by UVCS
onboard SOHO. They found that at a given time the line-of-sight velocity
within the CMEs increased with distance from 1.5 to 3~R$_S$. We suspect
therefore that the conventional leading edge velocity may not well represent
the characteristic speed of the entire CME mass and that conventional kinetic
energy estimates may be too high.
However, \citet{Ventura:etal:2002} could only discriminate the line-of-sight
velocity which for a CME propagating close to the plane of the sky
forms only a small component of the total velocity.

In this paper we propose a new CME analysis which allows a more refined
determination of the CME kinematics. Instead of integrating the
column density visible in successive, background-cleaned coronagraph
images to monitor the entire visible CME mass, we only integrate
azimuthally, normal to the dominant propagation direction.
This yields radial mass density profiles at any observational time which allow us to
follow the radial CME density transport in much more detail.
Given that this mass density is propagated according to a
continuum equation, we can derive the radial velocity
profiles of this propagation.

We demonstrate the new approach by applying it to an observed CME.
The according data is presented in section 2, while the method is outlined
in detail in section 3 and applied in the following sections. In the 
last section are conclusions and discussions.

\section{Data}

\begin{figure} 
\centering
\includegraphics[width=16.cm]{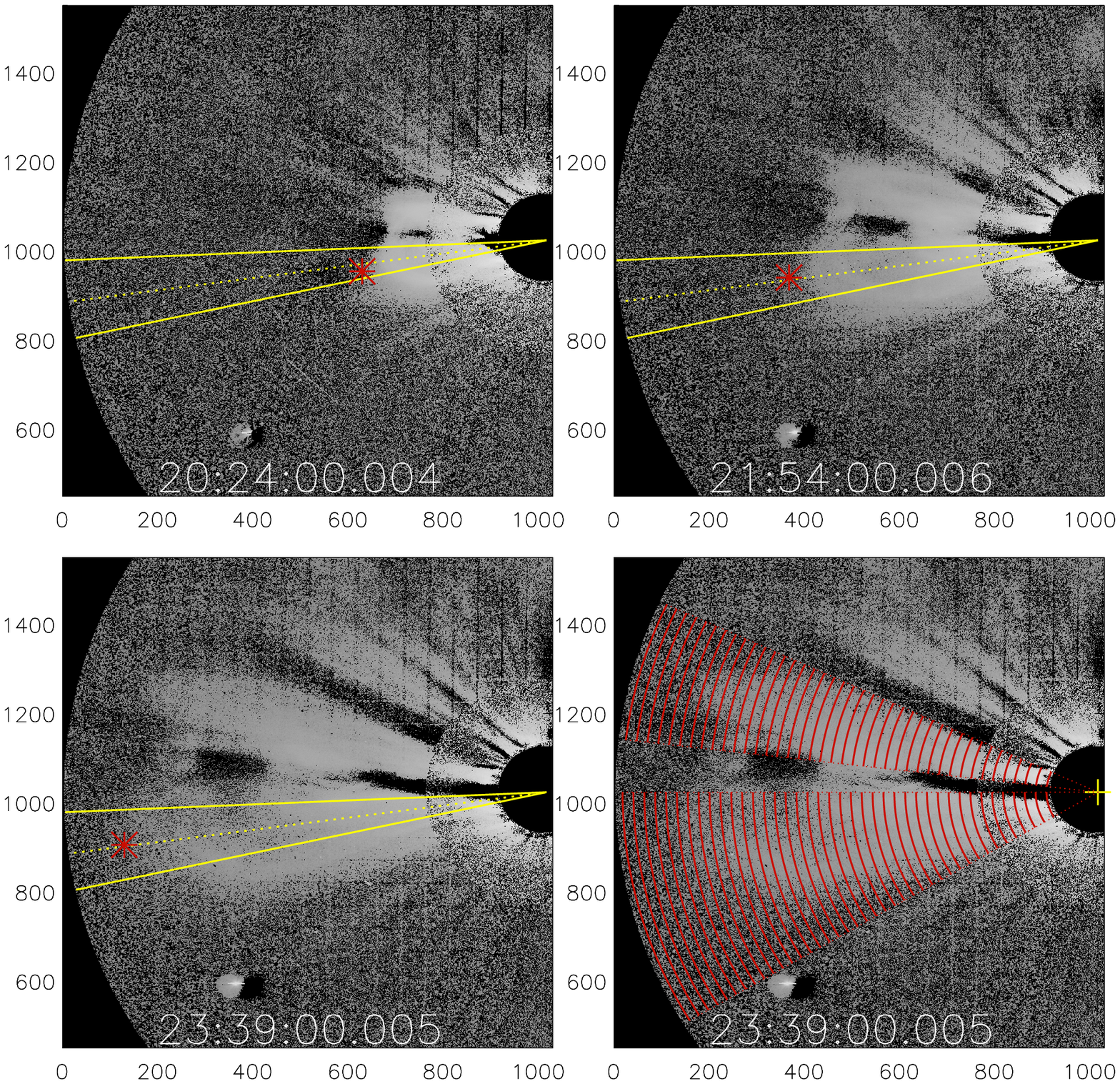}
\caption{Examples of combined COR1 and COR2 images for three different observational times.
The units are in pixel. The red asterisks mark the mean position
of the leading edge from eight independent hand-traced positions for each time
frame. All leading edge positions fall into the solid yellow cone of 10-degree width.
In the bottom right panel, the red circular sections display the boundaries of the 39
shells inside which the mass is integrated.}
\label{fig:cor2}
\end{figure}

A filament associated with a CME erupted at around 17:30~UT on
November 10 2010 from an active region at Stoneyhurst
coordinates 27$^\circ$E, 10$^\circ$N. The CME was then
observed in the field of 
view of the coronagraphs onboard the
\textit{Solar TErrestrial RElations Observatory} (STEREO), and the
coronagraphs onboard the \textit{Solar and Heliospheric Observatory} (SOHO).
In Figure~\ref{fig:cor2}, we present the combined COR1 and COR2 images at
three selected time instances taken by STEREO A. In each case, pre-event
background images were subtracted.
Figure~\ref{fig:cor2} therefore essentially shows the excess brightness 
produced by the CME. The dark region around the central latitude of
10$^{^\circ}$N is due to the presence of a streamer in the pre-event
image which dissolved after the eruption.

\begin{figure} 
\centering
\includegraphics[width=16.cm]{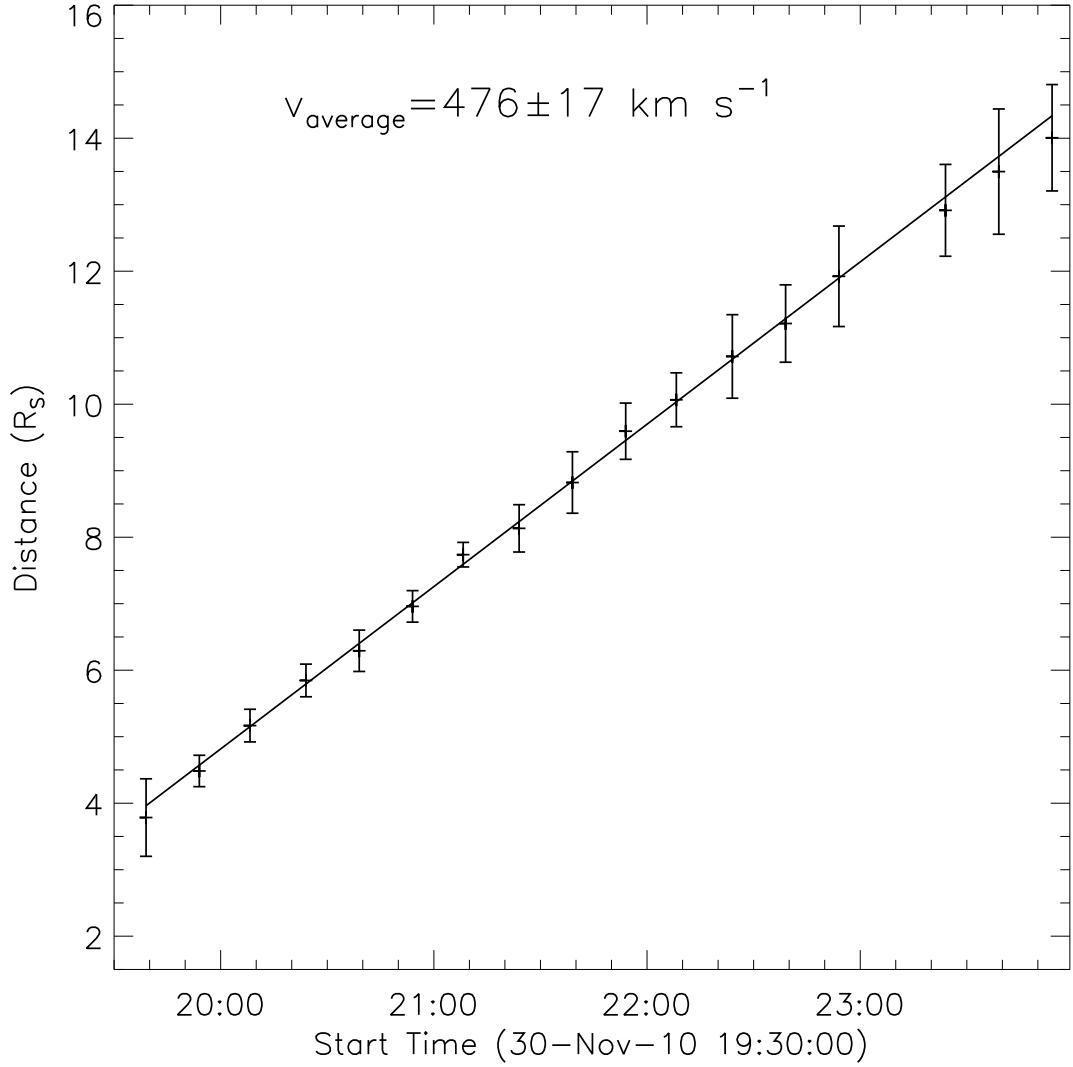}
\caption{The distance-time plot of the leading edge and a corresponding linear fit.
The symbols represent the mean and the $3\sigma$ uncertainties of the 
leading edge distance.
An averaged speed and its $3\sigma$ uncertainty derived from the linear fit is 
indicated in the upper part. }
\label{fig:leading_edge}
\end{figure}

For later comparison, we hand-traced the leading edge position of
the CME in a conventional way \citep[e.g.,][]{Bein:etal:2011}. As the leading edge,
we define the most distant point of the CME cloud boundary. This boundary 
cannot always be located precisely because the CME brightness rises from
a varying noise level at different distances from the center of the Sun. In order
to determine the uncertainty, we repeated the tracking independently for
eight times for each frame, and derived the mean and its variance of the
leading edge position for each time frame. The mean leading edge positions
were all located inside the yellow cone superposed in Figure~\ref{fig:cor2}.
The averaged leading edge positions are marked by red asterisks. The leading 
edge distance versus time and its associated $3\sigma$ uncertainty are shown in
Figure~\ref{fig:leading_edge}. The average speed of about 476~$\mathrm{km s^{-1}}$
yields a fairly close linear fit as shown by the straight line.

The simultaneous views of the CME from different perspectives provide a good
constraint to determine its propagation direction in three dimensions. On
November 10 2010, the two STEREO spacecraft were separated from the Earth by
about 85$^\circ$ each. Applying the graduated cylindrical shell forward
modeling of \citet{Thernisien:etal:2009}, we derived a propagation longitude
about 27$^\circ$E from the Sun-Earth line. This complies well
with the location of the source region and implies a propagation at $\alpha=
22^\circ$ behind the eastern plane-of-the sky of STEREO A.

\section{Method}

\subsection{Mass estimate}

A great advantage in the interpretation of white-light coronagraph images
is the fact that the major part of its brightness signal is produced
by Thomson scattering which depends linearly in the density $N_e$ of the
scattering free electrons.
The major non-Thomson scattered contributions like stray light and
dust scatter are removed by subtracting
the pre-event background image.
The remaining image brightness then should show the CME density
distribution produced by Thomson scattering alone.
The details of this scattering process have
already been developed more than fifty years and yield for the
total image brightness in a given image pixel
\citep{Minnaert:1930,vandeHulst:1950,Billings:1966}. 
\begin{gather}
B=B_\odot\frac{\pi\sigma_e}{2}
   \int_\mathrm{LOS} F(r,\chi, u) \; N_e(\mathbf{r}) \;d\ell,
\label{equ:totB}\\
    F(r,\chi,u)=2\frac{(1-u)C(r)+uD(r)}{1-u/3}
               - \frac{(1-u)A(r)+uB(r)}{1-u/3}\sin^2\chi
 \nonumber
\end{gather}
where
$B_\odot$ is the mean solar brightness (MSB) of the solar disk,
$\sigma_e$ is the coefficient of the differential Thomson scattering
cross section and the parameter $u$ accounts for the solar disk limb
darkening.
The variables $A,B,C$ and $D$ are known functions (van de Hulst coefficients)
of the distance $r$ of the scattering location from the solar center and
$\chi$ is the respective scattering angle for a photon from the solar center
to the observer.
The integration has to be performed along the line-of-sight (LOS)
defined by the effective view direction of the image pixel and
$s$ is the geometrical distance along the LOS.
We label the different LOSs by $\rho$, their closest distance to the Sun
and set $s=0$ at this closest approach to Sun so that
$r=\sqrt{\rho^2+s^2}$.

Since we do not know the true variation of $N_e$ along the LOS, Equation~
\ref{equ:totB} cannot be evaluated without further assumptions. We adopt the
approach proposed by \citet{Colaninno:Vourlidas:2009,Carley:etal:2012} which
assumes the density is concentrated entirely in the mean propagation plane at
an angle $\alpha$ off the plane-of-the-sky. Formally, this results in replacing
$F(r,\chi,u)$ in (\ref{equ:totB}) by $F(\rho/\cos\alpha,\pi/2-\alpha,u)$.
For the limb darkening coefficient $u$ we adopt a value of 0.56.

\begin{figure} 
\centering
\includegraphics[width=16cm]{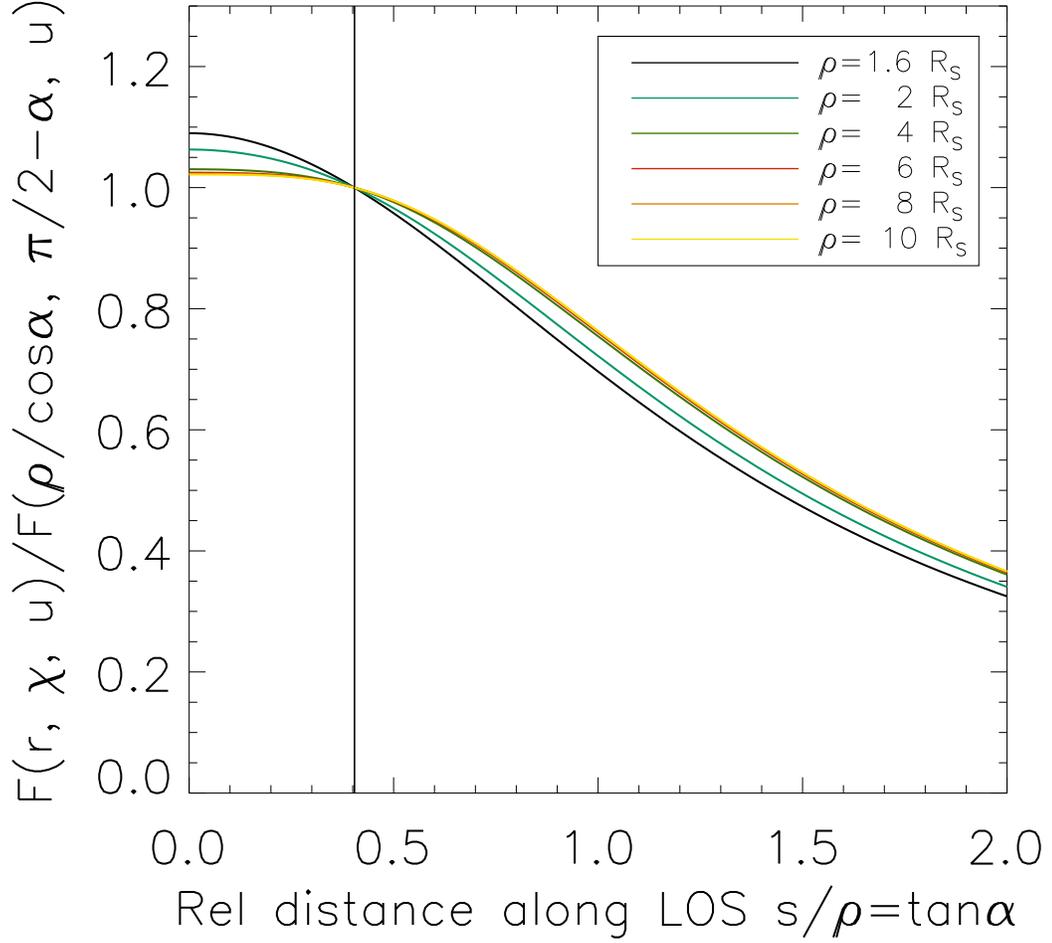}
\caption{Thomson scattering weight function $F(r,\chi,u)$ along the
 line-of-sight distance $s$ for various LOS paths with distance of 
 $\rho=1.6,2,4,6,8, 10~R_S $ from the center of the Sun. 
 The ordinate $F$ is normalized to the respective value at $s=\rho\tan\alpha$ where $\alpha$
 is the angle of $22^\circ$ between the CME propagation plane and the POS.
 The abscissa is normalized by $\rho$. The vertical line represents $\alpha=22^{\circ}$.}
\label{fig:ScattWeight}
\end{figure}

In Figure~\ref{fig:ScattWeight} we show the variation of the weighting
coefficient $F$ along a LOS at different distances $\rho$ from the solar
center normalized to its value on the propagation plane at angle $\alpha$.
If we knew the shape of the density distribution along the LOS at
some distance $\rho$, the true column density $\int N_e\;d\ell$ differs
from the approximate $\int\tilde{N}_e\;d\ell$ derived above by a factor
obtained from the convolution of the known normalized density
distribution along the LOS with the respective normalized weighting
coefficient in Figure~\ref{fig:ScattWeight}. The all-in-propagation-plane
assumption yields
\begin{gather}
   \int_\mathrm{LOS} F(r,\chi, u) \; N_e(\mathbf{r}) \;d\ell
=  F(\frac{\rho}{\cos\alpha},\frac{\pi}{2}-\alpha,u)
   \int_\mathrm{LOS} \tilde{N}_e(\mathbf{r}) \;d\ell
\nonumber\\ \text{or}\qquad
 \frac{\int_\mathrm{LOS} N_e(\mathbf{r}) \;d\ell}
      {\int_\mathrm{LOS} \tilde{N}_e(\mathbf{r}) \;d\ell}
=[ \int_\mathrm{LOS} \frac{F(r,\chi, u)}
              {F(\frac{\rho}{\cos\alpha},\frac{\pi}{2}-\alpha,u)} \;
  \frac{N_e(\mathbf{r})}{\int_\mathrm{LOS} N_e(\mathbf{r}) \;d\ell'}
  \;d\ell]^{-1}
\label{InPlaneAss}
\end{gather}
We note that normalized weighting coefficient does hardly depend on the
distance $\rho$ of the LOS from the solar center. I.e., a CME plasma volume
propagating radially with a constant angle of propagation off the POS, i.e.
with a constant $s/\rho$, receives the same LOS weighting. For an extended CME
with an unknown LOS distribution, the correction we have to apply to the
column density determined by the all-in-propagation-plane assumption does
therefore hardly depend on $\rho$ if the CME propagates strictly radially.

We further note that a closer inspection shows that the $s$-dependence of the
weighting coefficient in Figure~\ref{fig:ScattWeight} is due to the approximate
$1/r^2$ dependence of the van de Hulst coefficients which reflects the
decrease of the incident Sun light with distance from the solar center. The
influence of the variation with scattering angle through the differential
Thomson cross section is only minor.

\subsection{Mass profiles and their uncertainty analysis}

After the conversion of the excess brightness into the column electron
density per pixel in the CME region, we can estimate the according column
mass assuming a He$^{2+}$/H$^+$ composition of 10\%.
To analyze the mass transport inside the CME, we integrate the CME column
mass density from each pixel over discrete shells of radius $r$ centered at Solar
center to obtain mass $m(r,t)$ per unit length in radial direction at the 
times $t$ of the respective image. The discretization of radial shells is indicated in the
bottom right panel of Figure~\ref{fig:cor2}. To define 
shells, we have divided the radial range from 1.6 to 14.8~R$_S$ into 39 equal intervals 
and each 0.34 R$_S$ wide. This discretization is large enough to
raise the integrated mass in each shell significantly above the noise level, 
and small enough to sufficiently resolve the radial mass density distribution.
In order to reduce the noise in the estimates of $m$ we limit the integration
in latitude to the smallest possible sector which includes the CME. Note that
we also excluded the narrow wedge near the center of the CME where the
pre-event streamer disturbs the excess brightness.

\begin{figure} 
\centering
\includegraphics[width=16.cm]{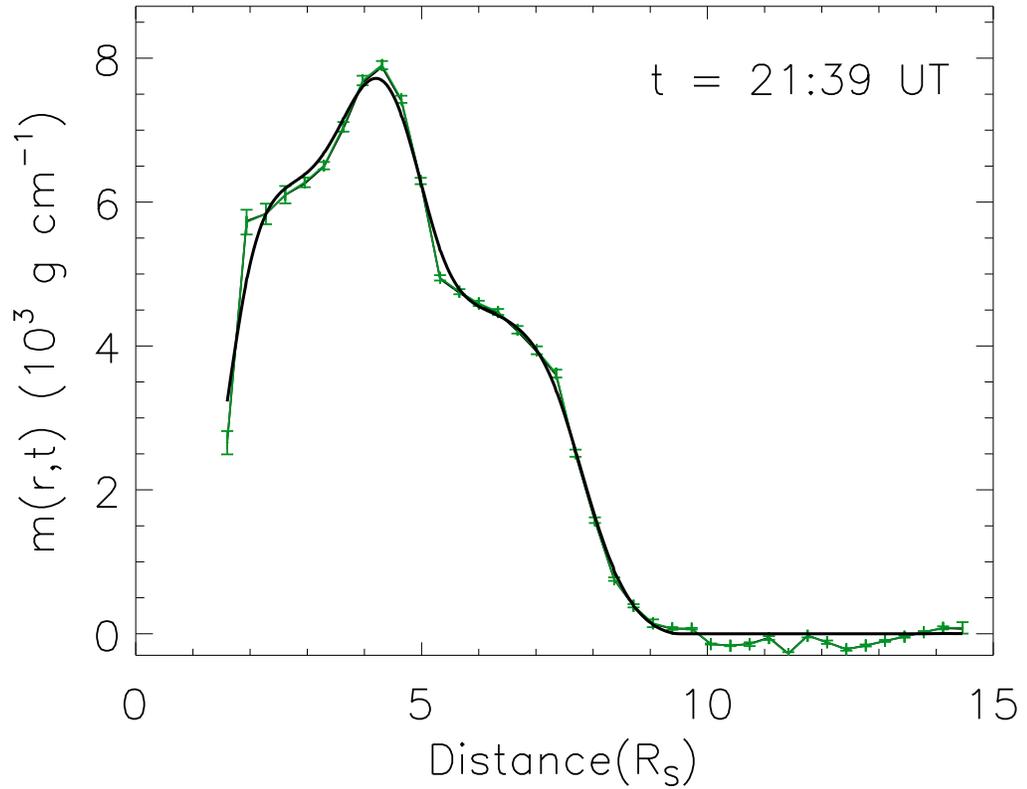}
\caption{Example of the radial mass density profile at 21:39~UT. The green curve and the respective curve
symbols were obtained from integrations of the image brightness. The error bars represent the $3\sigma$ of 
the integration results obtained for different selections of the CME sector and propagation angle. 
The black curve is a smoothed fit obtained from a smoothing spline.}
\label{fig:mass_err_smooth}
\end{figure}

There are two possible sources of systematic 
error in $m(r,t)$. One error in $m(r,t)$ may be due to the assumption 
in Equation~\ref{InPlaneAss} about the line-of-sight distribution of the 
CME mass, in particular due to a wrong longitudinal propagational angle 
$\alpha$. A second error in $m(r,t)$ may have been introduced by the somewhat 
arbitrary selection of the latitudinal
integration boundaries. Choosing them too wide causes additional noise to
be added to $m(r,t)$.
To estimate these possible errors, we repeated the calculation of $m(r,t)$ twelve
times for slightly different integration boundaries and also varied
the propagation angle $\alpha$ within $22\pm5^\circ$. As an example, in 
Figure~\ref{fig:mass_err_smooth} the average and $3\sigma$ uncertainty
of the radial mass density profile from its twelve integrations at 21:39~UT
is displayed in green.

\begin{figure} 
\centering
\includegraphics[width=16.cm]{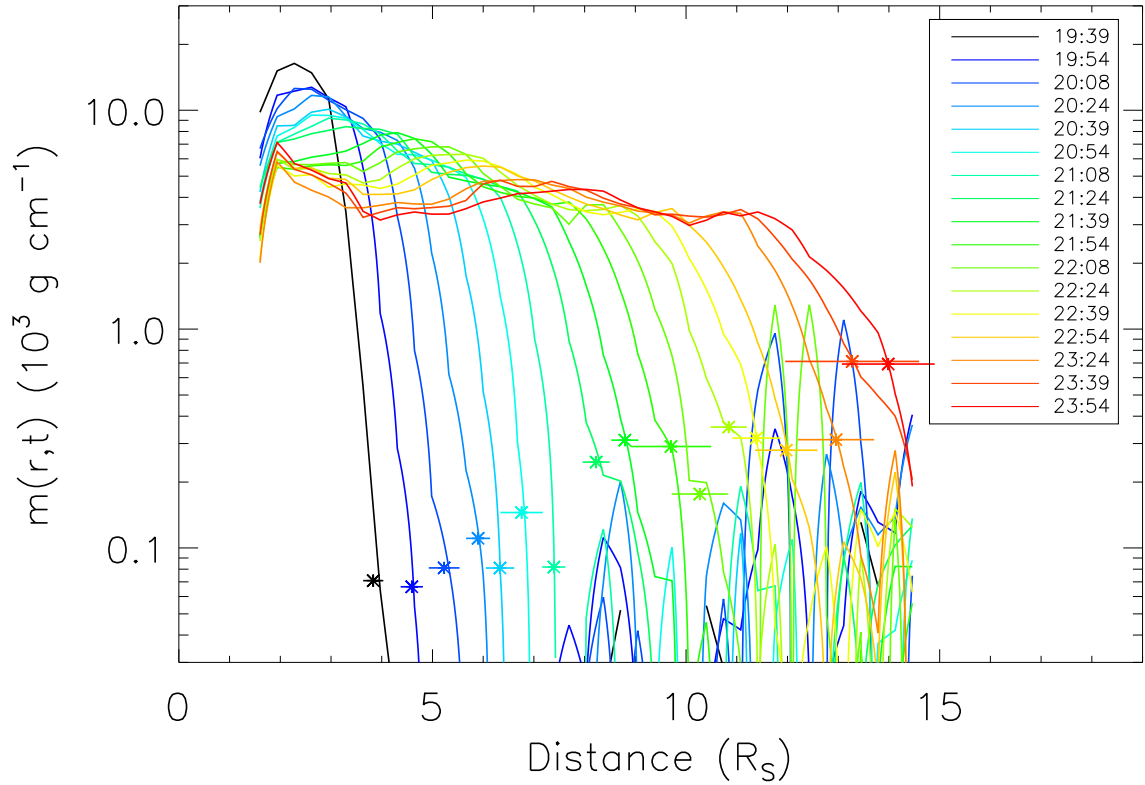}
\caption{Time series of the mass profiles as a function of distance plotted in 
logarithm scale. The lines in different colors represent the 
mass profiles at different times. The color and time correspondence is shown
in the upper right legend. The positions of the leading edge and their 
associated uncertainties are indicated by asterisks and errorbars.}
\label{fig:mass_pf_log}
\end{figure}

Figure~\ref{fig:mass_pf_log} displays the whole time series of mass 
density profiles $m(r,t)$ thus obtained. The colors from black to red indicate time $t$
in increasing order. On each curve, the asterisk marks the position of the
hand-traced leading edge. We can see that the leading edge corresponds to
finite, different mass densities which slightly increase with distance. 
There is still some residual CME mass ahead of the leading edge.
The spikes in $m(r,t)$ at about $r>10 R_S$ are due to the increasing
noise in the coronagraph images towards the outer boundary of
the field of view.

In order to derive other quantities from $m(r,t)$ and their uncertainty,
we first fitted smoothing splines to $m(r,t)$ obtained so far.
The amount of smoothing was controlled by the variance of the individual
profiles from their average. One such smoothing spline is shown in black in 
Figure~\ref{fig:mass_err_smooth}.
Note that well ahead of the CME, $m(r,t)$ was set to zero
before the smoothing. The mass measurements start to be sufficiently 
reliable at $\simeq2~R_S$, about 0.5~$R_S$ above the COR 1 occulter.

\subsection{Flow speed derivation}
\label{sec:SpeedDerv}

If we assume that there is no mass contribution from the pileup of the 
solar wind around the CME, the mass flow is described by a continuity
equation
\begin{equation}
\frac{\partial{\rho}}{\partial{t}} + \boldsymbol{\nabla}\cdot(\rho\mathbf{v})=0.
\label{eq:conti3D}\end{equation}
The radial mass density profiles $m(r,t)$ which we have derived 
above are integrals of
the volume mass density $\rho$ over cylinder surface sections which are spanned
by the weighted LOS integration in one direction and heliographic latitude
section in the other.
Clearly, if $\mathbf{v}$ has only a radial component, the
$\boldsymbol{\nabla}\cdot$-operation commutes with both integrations and
Equation~\ref{eq:conti3D} could be transformed to
\begin{equation}
\frac{\partial{m}}{\partial{t}} + \frac{\partial}{\partial r}(m v_r)=0
\label{eq:conti1D_Euler}\end{equation}
We expect that the major velocity component of the CME is indeed radial
so that Equation~\ref{eq:conti1D_Euler} gives a good approximation to the evolution
of the radial mass density $m$.

It is obvious that mass flow in latitudinal direction can be coped with by
partial integration and does not change Equation~\ref{eq:conti1D_Euler} as long as the
longitudinal integration boundaries were chosen wide enough to include the
entire visible CME signal. The effect of a latitudinal component of the mass
flow is more difficult to estimate. From Figure~\ref{fig:ScattWeight} we see
immediately that a change of the propagation angle $\alpha$ or of the LOS
density profile in the CME with distance $\rho$ could lead to a brightness
change without changing the true column mass density. However, from many CME
observations it has become evident that the non-radial components of the CME
propagation is small. We will therefore in the following drop the index $r$ of
$v_r$ in Equation~\ref{eq:conti1D_Euler} and assume this equation to be an adequate
description for the evolution of $m$, neglecting possible changes in the
mass density caused by the non-radial components of $\mathbf{v}$.

At first sight, Equation~\ref{eq:conti1D_Euler} seems of little value as long as
$v(r,t)$ is not known. However since $m(r,t)$ is measured, it opens a chance
to derive the velocity. Given that $m(r\rightarrow\infty,t)$ vanishes, an
integration of Equation~\ref{eq:conti1D_Euler} over radius with variable lower bound
yields
\begin{equation}
   v(r,t)=-\frac{1}{m(r,t)} \;\frac{\partial}{\partial{t}}
            \int_r^\infty m(r',t) dr'
\label{eq:contiV}\end{equation}
That is to say, at a given time, the total mass change beyond $r$
comes from the mass flow through the shell at $r$.

\begin{figure} 
\centering
\includegraphics[width=16.cm]{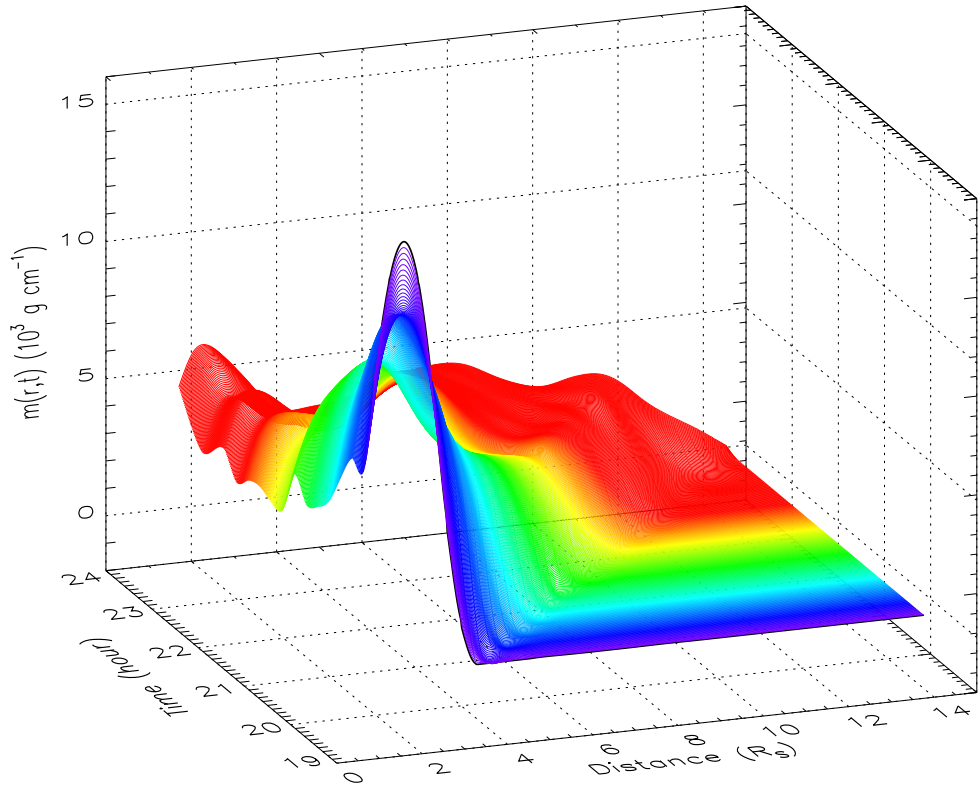}
\caption{2D cubic spline interpolation of mass profiles from 19:39 
to 23:54~UT and between 1.6 and 14.8~R$_S$. The $X$ axis indicates
the distance, and $Y$ axis indicates the time. The first mass profile
at 19:39~UT is displayed in black, and the following mass profiles change
their color with time from violet to red.}
\label{fig:mass_prof_interp}
\end{figure}

\begin{figure} 
\centering
\includegraphics[width=15.cm]{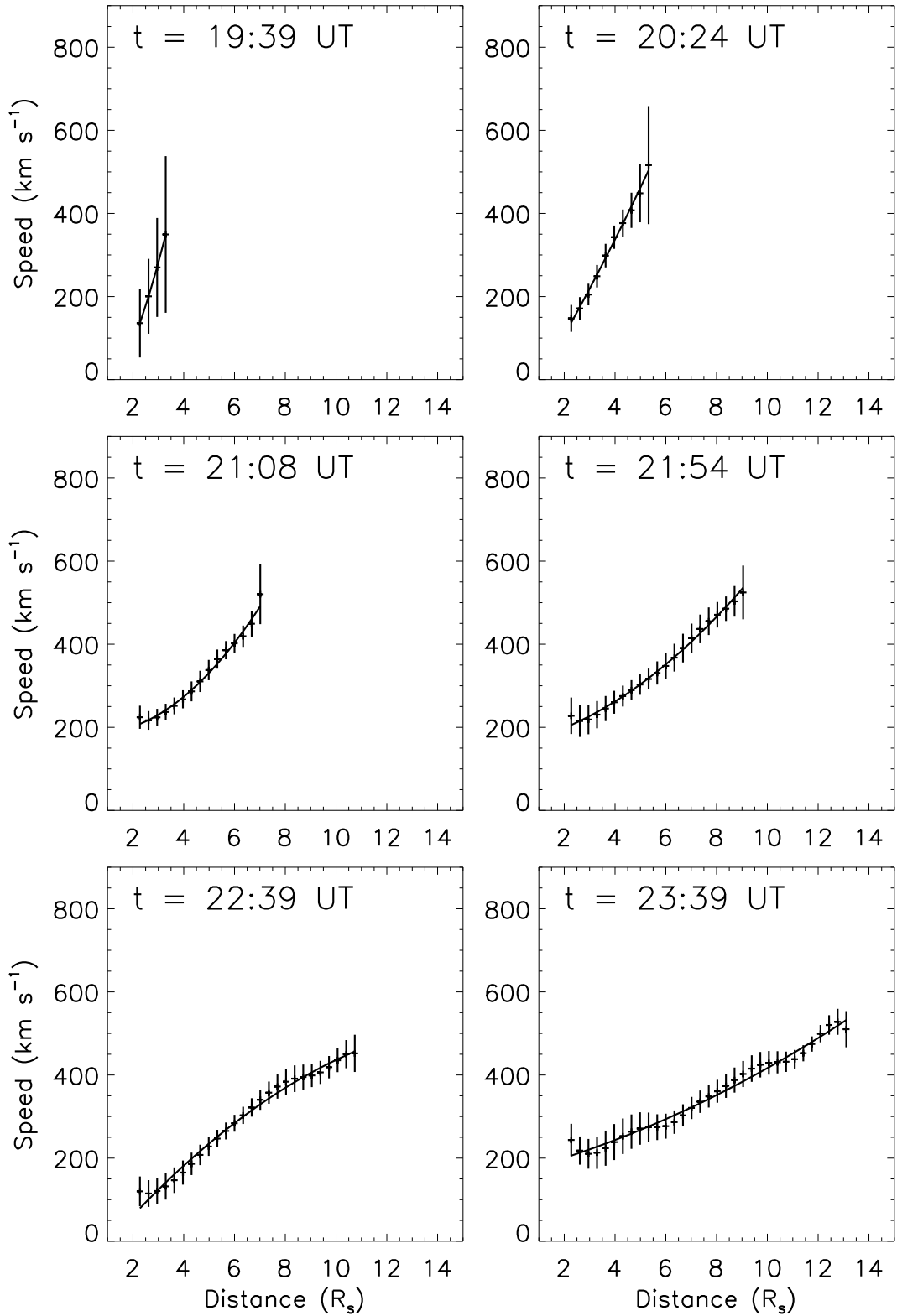}
\caption{The distribution of flow speed with distance for six selected time
instances. The error bars in each panel were derived from the uncertainties of the mass profiles.}
\label{fig:v_r_tj}
\end{figure}

\begin{figure} 
\centering
\includegraphics[width=12.cm]{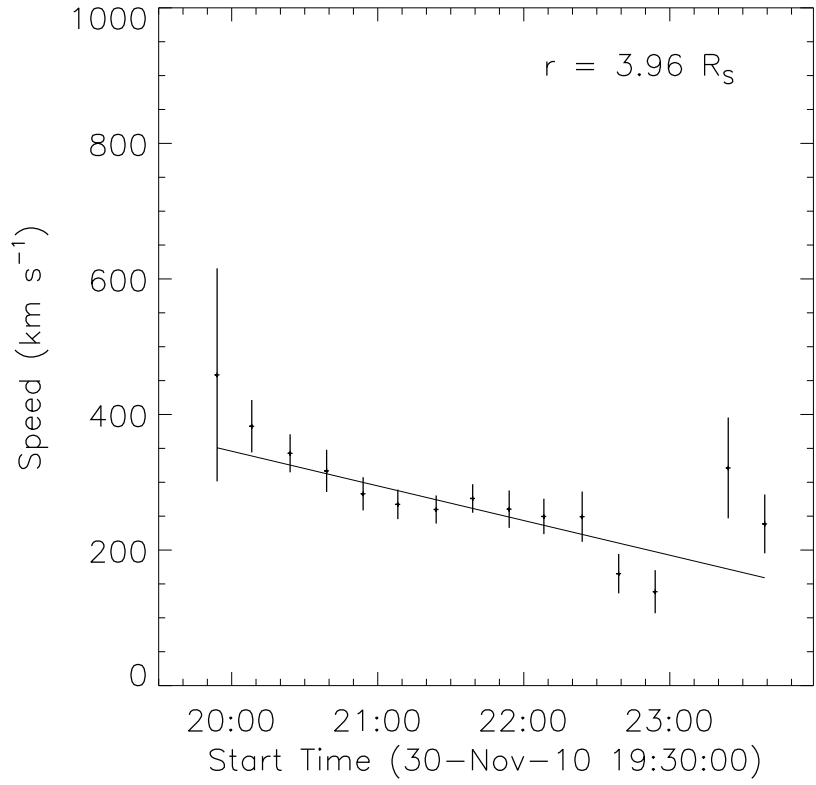}
\caption{An example of the evolution of flow speed with time at a distance of $r=3.96~R_S$.
The error bars are the same as in Figure~\ref{fig:v_r_tj}.}
\label{fig:v_ri_t}
\end{figure}

In Equation~\ref{eq:contiV}, the calculation of the velocity 
requires the time derivatives of mass profiles. To reliably calculate 
these derivatives, we interpolated the mass profiles using cubic splines.
The result of such interpolations
is presented in Figure~\ref{fig:mass_prof_interp}. The first mass profile
at 19:39~UT is displayed in black, and the following mass profiles change color with 
time from violet to red. The statistical error in 
$v(r,t)$ was obtained by repeating the
calculation 300 times with mass density profiles $m(r,t)$ contaminated
by random noise profiles of an amplitude which agrees to the
standard deviation determined for $m(r,t)$. The variation in the resulting
velocities was used to determine the error in $v(r,t)$. In such
a way, we can derive an Eularian flow field within the CME at each 
interpolated space-time position. Some of the velocity profiles and associated 
errorbars at defined shell positions and at selected representative times  
are displayed in Figure~\ref{fig:v_r_tj}. The solid curve in each panel is a 
quadratic fit to the derived velocity profile. Figure~\ref{fig:v_ri_t} shows
an example of the time evolution of the Eulerian velocity at a given distance. 
We find that the Eulerian velocity decreases with time at all distances.

So far, we have ignored possible mass sources in the CME continuum equation 
\ref{eq:conti3D} and \ref{eq:conti1D_Euler}. One such possibility which
has been discussed is the pile up of solar wind mass in front of a fast
propagating CME. We see similar density enhancements ahead of fast solar wind
streams where they run into the warped up current sheet. Formally, this could
be taken into account by adding a source term in Equation~\ref{eq:conti1D_Euler}. The
effect on the velocity derived can be seen straight forwardly. If a source is
present, then
\begin{gather*}
    \frac{\partial}{\partial t}m
  =-\frac{\partial}{\partial r}(v m) + s(r,t)\\
\text{Set}\qquad f(r,t)=\int^r s(r',t) dr' \quad \text{then} \\
    \frac{\partial}{\partial t}m
  =-\frac{\partial}{\partial r}(v m) + \frac{\partial}{\partial r} f
  =-\frac{\partial}{\partial r}((\underbrace{v-f/m}_{v'}) m)
\end{gather*}
Hence, the source term is absorbed in the flow velocity which we obtain
from Equation~\ref{eq:contiV}.
For example if there is mass pile-up in front of the CME, $f$ should grow near
the front side of the CME from zero to the piled-up mass well ahead of the
CME. This should lead to a strong decrease of the velocity $v'$ since, at the
front side of the CME, $m$ drops with distance and the velocity contamination
$f/m$ increases even stronger than $f$ alone. Therefore a mass pile-up should
show as a decrease in the velocity profile near the CME front. In
Figure~\ref{fig:v_r_tj} we see that the velocity at the CME front is
persistently about 450-500 km s$^{-1}$. Note that the error in the most
distant velocity estimate has a relatively large uncertainty because it was
obtained in Equation~\ref{eq:contiV} from a division by a small mass density
$m(r,t)$.

\subsection{Validation of flow speeds}

To check the consistency of the velocity $v(r,t)$ derived above we have
used it in the continuity equation, i.e., Equation~\ref{eq:conti1D_Euler} to 
predict the radial
mass density profile $m(r,t_j)$ from a given initial value $m(r,t_{j-1})$.
Both times $t_{j-1}$ and $t_{j}$ were chosen to coincide with the observational times
the coronagraph images were taken so that the respective mass
density profiles were those obtained directly from the integration of
image data. The continuity equation was integrated by a straight forward
Lax-Wendroff scheme. The boundary condition
varies linearly from $m(r_0, t_{j-1})$ to $m(r_0, t_j)$ at each time step.

\begin{figure} 
\centering
\includegraphics[width=16.cm]{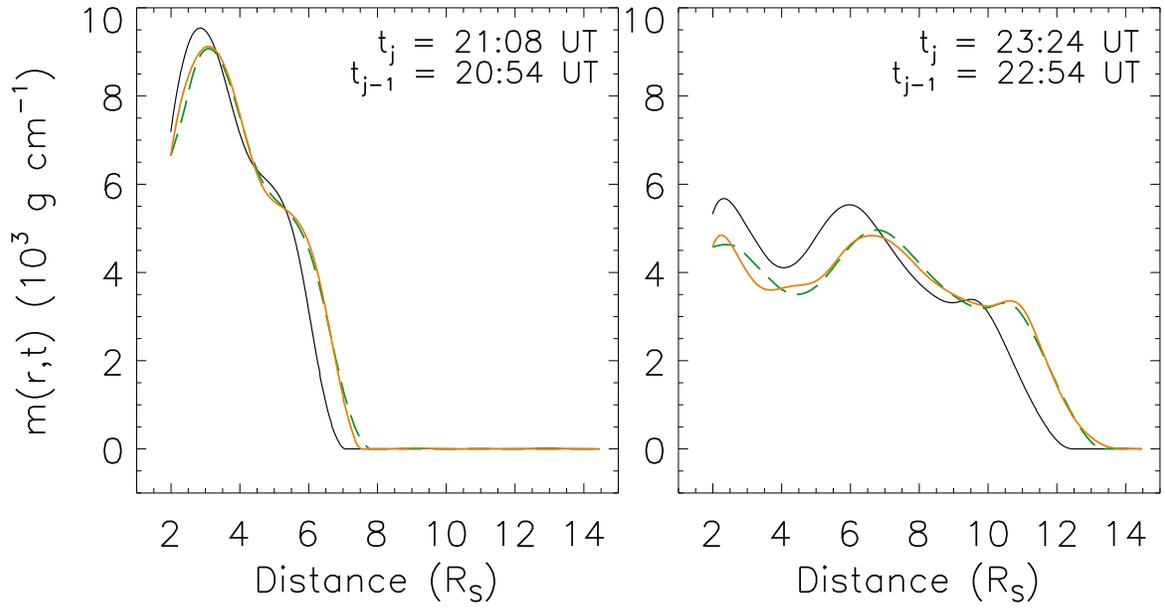}
\caption{Comparison of measured and calculated mass profiles.
The solid lines in black and orange show the measured $m(r,t)$ at times $t_{j-1}$ and $t_j$.  
The dashed line in green is the $m(r,t)$ at $t_j$ calculated from the profile at $t_{j-1}$.}
\label{fig:v_validate}
\end{figure}

Figure~\ref{fig:v_validate} shows examples at two pairs $t_{j-1}$ and
$t_j$ of successive image times.
The solid black and orange lines are the measured mass density
profiles at $t_{j-1}$ and $t_j$, respectively. The dashed green line is
the profile calculated from $t_{j-1}$ to $t_j$ using the flow speeds
$v(r,t)$ derived in the previous subsection. It can be seen that
the predicted and measured mass density profiles fit fairly well.

\section{Results}

The knowledge of the velocity and mass distribution allows us to study the CME
kinetics in much more detail than was possible in many previous
investigations. However, $v(r,t)$ is the velocity is in the Eulerian frame
with $r$ and $t$ considered independent variables which is sometimes difficult
to interpret. It is more instructive to see the CME mass transport in the
frame of the individual CME mass element.

\subsection{Lagrangian trajectories and leading edge positions}

A change of Equation~\ref{eq:conti1D_Euler} to the Lagrangian frame is achieved first
by integrating the velocity to the Lagrangian or material path of a mass
element
\begin{equation}
  \frac{\partial}{\partial t} R(t,t_0)=v(r=R(t,t_0),t)\quad
  \text{with initial condition}\quad R(t_0,t_0)=r_0
\label{eq:conti1D_lagrangePath}\end{equation}
In this formulation, the paths $R$ are labeled according to the time
$t_0$ (2nd argument) when they pass the distance $r_0\simeq 2\, R_\odot$
where we start our velocity derivation.

Along these paths the advected mass density $m(r=R(t,t_0),t)$
then changes in time according to
\begin{equation}
\frac{dm}{dt} = -m\frac{\partial v}{\partial r}(r=R(t,t_0),t)
\label{eq:conti1D_lagrangeMass}\end{equation}
The error bars assigned to each path represent the estimated uncertainty
obtained again by repeated integrations of Equation~\ref{eq:conti1D_lagrangePath} of
the average flow speed perturbed with random noise profiles of amplitude in
agreement with the standard deviation of $v(r,t)$.

\begin{figure} 
\centering
\includegraphics[width=16.cm]{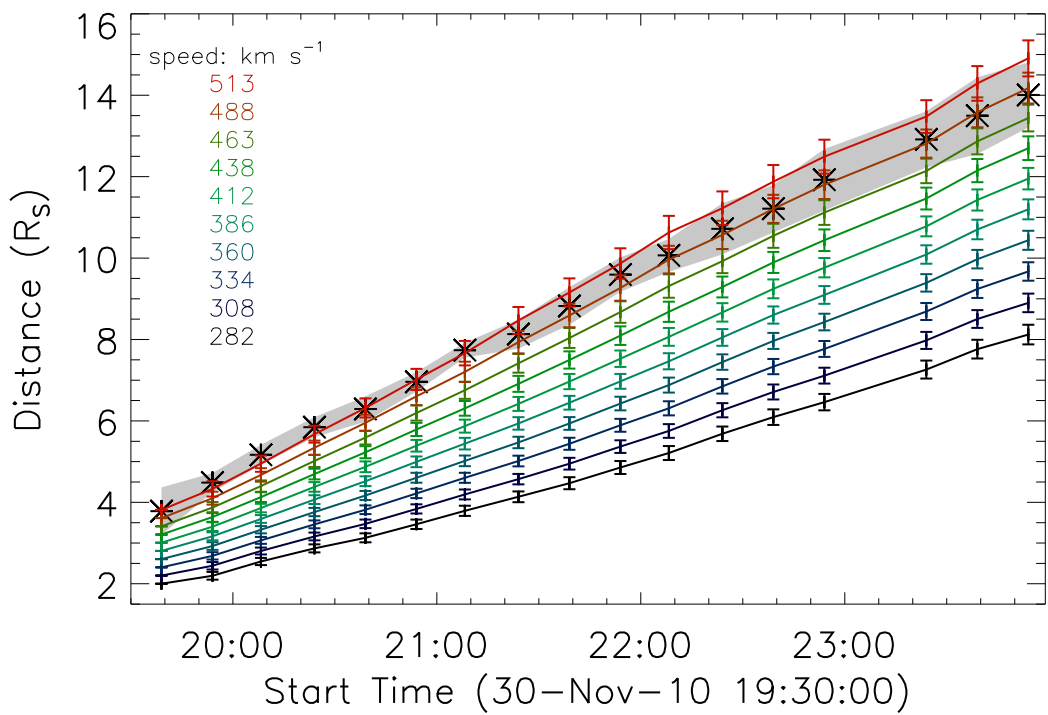}
\caption{Some Lagrangian trajectories derived from the Eularian 
velocity profiles. At their starting time 19:39~UT, they were equally 
distributed within the CME region between 2 and 3.8~$R_S$. The error bars are derived from the 
uncertainties of the velocity profiles. The asterisks and the shaded 
gray color mark the leading edge positions and their associated uncertainties.
The corresponding averaged Lagrangian speeds are indicated in the left in the 
same color as the respective trajectories.}
\label{fig:lagran}
\end{figure}

From the derived Eularian velocity field in the last section, 
it is possible to derive the Lagrangian path of a mass element starting from
$r_0$ at time $t_0$. In Figure~\ref{fig:lagran}, we present the Lagrangian
path of 10 mass elements starting at 19:39~UT. At 19:39~UT, the CME mass covers 
a range from 2 to 3.8~R$_{\odot}$. The selection of ten mass elements is to 
make the errorbars of the trajectories distinguishable between neighboring 
trajectories, and at the same time sufficiently resolve the properties 
of different trajectories.
For the CME studied here, each mass element propagates at almost a constant 
Lagrangian velocity. The height-time diagram of the Lagrangian paths in
Figure~\ref{fig:lagran} closely resembles a fan, however, each Lagrangian path
(and those in between) carrying a different amount of mass density.
If we approximate these velocities by exact constants, it is easily seen
that $\partial v/\partial r \propto 1/t$ and the mass density along a
Lagrangian path decreases as $\propto 1/R$.

Any significant derivation of the Lagrangian path from a straight line
indicates either an physical acceleration $dv/dt$ of CME masses or,
as discussed in section~\ref{sec:SpeedDerv}, the presence of a mass
source. In the example studied here we conclude from the constancy
of the Lagrangian trajectories that there is possibly no evidence for either
of the two.

We note that the asterisks, which represent the radial advance of the leading
edge are attached to the fastest Lagrangian path in the beginning but slightly
fall behind systematically at later times. The leading edge we have picked by
visual inspection (see Figure~\ref{fig:leading_edge}) may therefore not
exactly correspond to a material path as the name suggests. It seems that
as the CME front dims away at increasing distance from the Sun, there
is a tendency to place the leading edge position further inward where the
image brightness is more enhanced.

Another interesting aspect of Figure~\ref{fig:lagran} is the possibility to
extrapolate the Lagrangian paths backwards. We find that the fastest path with
a mean velocity of 521~$\mathrm{km\,s^{-1}}$ intersects the solar surface at 18:35 UT, about
one hour after a flare in the EUV images indicates the initiation of the
eruption at $t_\mathrm{flare}=17:30$ UT. The slower velocity paths extrapolate
to the surface at increasingly later times, the slowest velocity path with a
mean of 283~$\mathrm{km\,s^{-1}}$ intersects $r=1$ at 18:58 UT. A mass element experiencing
approximately a constant acceleration between some time $t_\mathrm{start}$ and
time $t_\mathrm{end}$ will propagate afterwards on a Lagrangian path which
intersects the solar surface at $(t_\mathrm{start}+t_\mathrm{end})/2$,
irrespective of the strength of the acceleration. If this simple model holds
true, the fast masses (red trajectories in Fig.~\ref{fig:lagran}) must have
been accelerated earlier than the slow masses (dark blue trajectories).

\subsection{CME energy} 

Usually the kinetic energy of a CME is estimated by simply using its total
mass and the leading edge speed. This ignores the extended distribution of CME
mass and velocity. For the CME we analyzed here, we see systematically
lower speeds with increasing distance from the leading edge.
Therefore, the conventional approach probably tends to overestimate the
kinetic energy of a CME.

\begin{figure} 
\centering
\includegraphics[width=14cm]{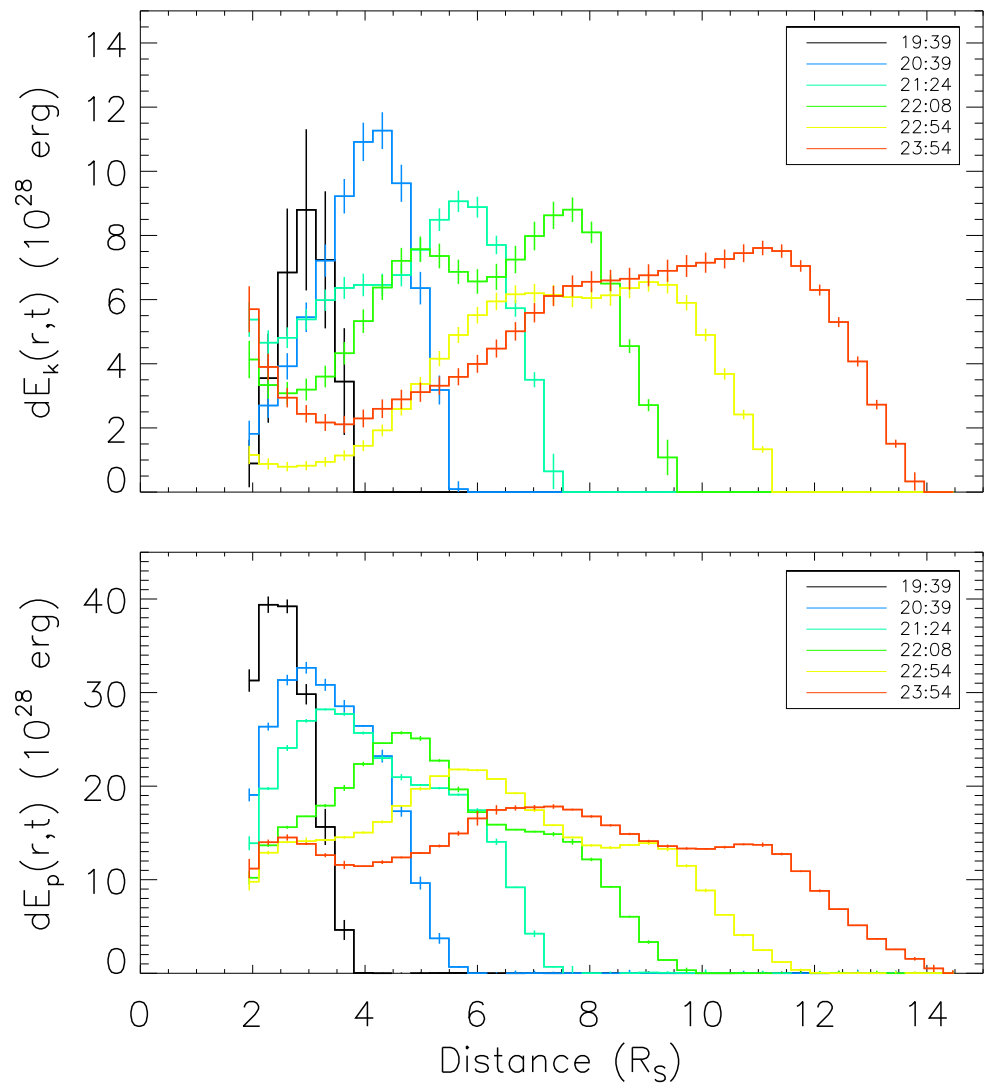}
\caption{Profiles of kinetic energy and gravitational potential energy in each 
shell at a few representative times. The energy profiles at
different colors represent the profiles at different times. The color-time
correspondence is shown in the upper right legend.}
\label{fig:Ek_Ep_prof}
\end{figure}

\begin{figure} 
\centering
\includegraphics[width=14cm]{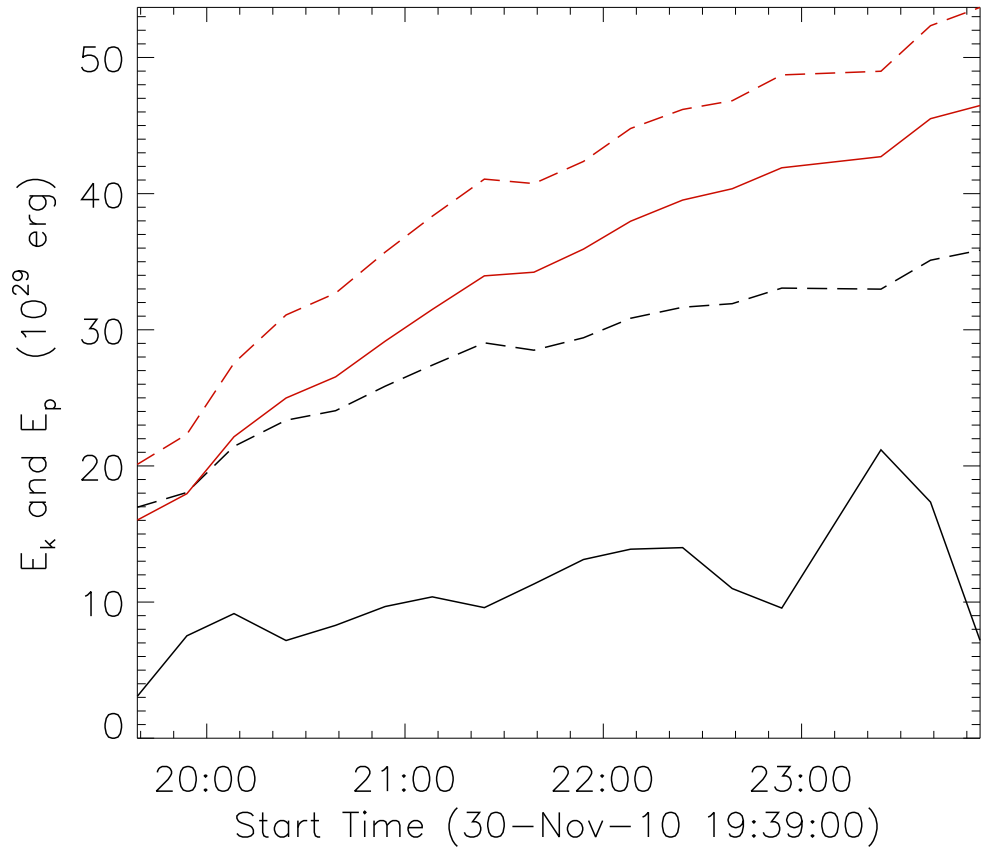}
\caption{Total kinetic energy and gravitational potential energy of the CME 
in our FOV are shown by black and red solid lines, respectively. The dashed lines 
represent the respective conventional energy estimates derived using the 
total mass and the leading edge motion.}
\label{fig:Ek_Ep_tot}
\end{figure}

In the upper panel of Figure~\ref{fig:Ek_Ep_prof}, we show the kinetic energy of the CME
between successive shells at $r_i$ and  $r_{i+1}$ at a few selected times
\begin{equation}
dE_\mathrm{k}(r_i,r_{i+1},t)=\int_{r_i}^{r_{i+1}} \frac{m(r,t)}{2}v^2(r,t)\;dr.
\label{equ:dE_k}
\end{equation}
The non-uniformity of the kinetic energy of CME shells can be clearly seen. 
In Figure~\ref{fig:Ek_Ep_tot} the time evolution of the total kinetic energies 
of the entire CME $E_k$ are marked in black. The values derived with the conventional 
method indicated by the dashed line can be four times larger than the values 
obtained with our more refined method indicated by the solid line. This large
discrepancy mainly comes from the $v^2$ term in Equation~\ref{equ:dE_k}.

Another energy we estimated is the gravitational potential 
energy above the solar surface according to:
\begin{equation}
dE_\mathrm{p}(r_i,r_{i+1},t)=\int_{r_i}^{r_{i+1}}m(r,t)g_\odot\frac{R_S}{r}(r-R_S)\;dr.
\label{equ:dE_p}
\end{equation}
The potential energy of each shell at the same selected times as the kinetic 
energy is illustrated in the lower panel of Figure~\ref{fig:Ek_Ep_prof}. The 
non-uniformity of the potential energy is also evident. The total potential energy
$E_p$ plotted in red in Figure~\ref{fig:Ek_Ep_tot} reveals that $E_p$ estimated
using total mass and leading edge distances may also overestimate the true value. 
The discrepancy between $E_p$ derived with these two different methods is much smaller, 
compared with that of $E_k$. 

\section{Conclusions and Discussions}

We have presented a new approach to analyze the mass evolution and advection
in a CME. It yields the radial mass density distribution and, as a derived quantity,
the radial velocity profile and its temporal evolution. This represents
much more detailed information than previous mass evolution
diagrams. As another advantage, our analysis is much less dependent
on subjective manual interference than the traditional tie-pointing
of the leading edge.

Our analysis allows us to detect acceleration also inside the CME and, at least
indirectly, from implausible velocity profiles to conclude about the existence
of CME mass sources in the solar wind, e.g., by pile-up. These conclusions
could not be drawn from the conventional total mass evolution plots. In the
present study of a slow CME, we did not find a speed decrease in the profiles
which could indicate such a pile-up expected ahead of CME. In future we will
apply our analysis to a fast CME where such a pile-up is more probable. In
this case, we will include a pile-up induced source term in the continuity
equation.

The velocity profiles obtained are compatible with a self-similar propagation
of the CME between 2 and 15 R$_S$, at least as far as the radial advection
of CME mass density is concerned. From the mass and velocity profiles inside
the CME we can determine where the most mass and kinetic energy is stored.
This could by forward extrapolation help to forecast effective travel times
which do not always seem to agree well with the speed of the leading edge. By a
careful backwards extrapolation towards the source it might in some cases
become possible to explore the initial distribution of the CME mass in the
corona. E.g. it could tell, how much of the CME mass originates in the
chromosphere or in protuberances and which portion is delivered by streamer
plasma at higher altitudes \citep{Kramar:etal:2009}.

It is also our aim to extend the profiles beyond the FOV
of COR2 and use the HI1 and HI2 coronagraph data of STEREO.
From in-situ observations of a solar wind transient, the speed at different
positions of an ICME can be measured directly. As an outlook, we plan to
compare the speed distribution in the FOV of COR to HI with the
mass flow observed in-situ at 1 AU.
Another extension of our study would be to apply it to the observations
from the two view directions of the STEREO spacecraft. The possible three
dimensional distributions of mass and velocity profiles would be
constrained by their independent measurements from two view points.

\acknowledgements
SOHO and STEREO are projects of international cooperation
between ESA and NASA. The SECCHI data are produced by
an international consortium of NRL, LMSAL, and NASA GSFC
(USA), RAL, and U. Birmingham (UK), MPS (Germany), CSL
(Belgium), IOTA, and IAS (France). SDO is a mission of NASA's
Living With a Star Program. L.F. and W.Q.G. are supported by
MSTC Program 2011CB811402, NSF of China under grants 11473070,
11233008, 11273065 and by the NSF of Jiangsu Province under grants BK2012889. L.F. also
acknowledges the Youth Innovation Promotion Association, CAS, for financial support. 
The work of B.I. is supported by DLR contract 50 OC 1301.


\end{document}